\documentclass[iop]{emulateapj}
\usepackage{graphics,graphicx}
\usepackage{subfigure}
\usepackage{color}
\begin{document}
\title{Fiber mode scrambler for the Subaru Infrared Doppler Instrument (IRD)}

\author{Masato Ishizuka$^{1}$,  Takayuki Kotani$^{2,3}$, Jun Nishikawa$^{2,3}$,Takashi Kurokawa$^{2,4}$, Takahiro Mori$^{4}$, Tsukasa Kokubo$^{4}$, and Motohide Tamura$^{1,2,3}$}
\affiliation{\\
$^{1}$Department of Astronomy, The University of Tokyo,  7-3-1, Hongo, Bunkyo-ku, Tokyo, 113-0033, Japan\\
\smallskip
$^{2}$National Astronomical Observatory of Japan, 2-21-1, Osawa, Mitaka, Tokyo, 181-8588, Japan\\
\smallskip
$^{3}$Astrobiology Center, 2-21-1, Osawa, Mitaka, Tokyo, 181-8588, Japan\\
\smallskip
$^{4}$Tokyo University of Agriculture and Technology, 3-8-1, Saiwai-cho, Fuchu, Tokyo, 183-0054, Japan}

\begin{abstract}
 We report the results of fiber mode scrambler experiments for the Infra-Red Doppler instrument (IRD) on the Subaru 8.2-m telescope. IRD is a fiber-fed, high precision radial velocity (RV) instrument to search for exoplanets around nearby M dwarfs at near-infrared wavelengths. It is a high-resolution spectrograph with an Echelle grating. The expected RV measurement precision is ${\rm \sim1\, m\,s^{-1}}$ with a state of the art laser frequency comb for the wavelength calibration. In IRD observations, one of the most significant instrumental noise is a change of  intensity distribution of multi-mode fiber exit, which degrades RV measurement precision. To stabilize the intensity distribution of fiber exit an introduction of fiber mode scrambler is mandatory. Several kinds of mode scramblers have been suggested in previous research, though it is necessary to determine the most appropriate mode scrambler system for IRD. Thus, we conducted systematic measurements of performance for a variety of mode scramblers, both static and dynamic. We tested various length multi-mode fibers, an octagonal fiber, a double fiber scrambler, and two kinds of dynamic scramblers, and their combinations. We report the performances of these mode scramblers and propose candidate mode scrambler systems for IRD.
\end{abstract}


\section{Introduction}
 Nearly 4000 exoplanets have been discovered, and Earth-like planets in the habitable zone (HZ, Huang 1959, Kopparapu et al. 2013) are drawing increasing interest, though extremely high precision is required to detect such planets. The radial velocity (RV) method is one of the most successful techniques for searching for exoplanets. A measurement precision for RV of ${\rm \sim10 \, cm\,s^{-1}}$ is required to detect Earth-like planets in the HZ around a Solar-type star, and this precision is far below the current achievable precision.
 
 M dwarfs are promising targets for the search for Earth-like planets, because their RV variation induced by planets is much larger than that of Sun-like stars. Thus Earth-like planets in the HZ can be detected with current RV measurement precision (${\rm \sim1 \, m\,s^{-1}}$). Since M dwarfs emit almost all their energy in the near-infrared (NIR) wavelengths, observation at NIR wavelengths is effective. 
 
We have developed the Infra-Red Doppler instrument (IRD) for the Subaru 8.2-m telescope. IRD is a fiber-fed spectrograph for use in the search for exoplanets around M dwarfs by a high precision RV method. Its wavelength coverage is 0.97--1.75 ${\rm \mu}$m with a spectral resolution of $\sim$70,000. A dedicated laser frequency comb has also been developed for precise wavelength calibration(Kokubo et al. 2016). The expected RV measurement precision is ${\rm 1\, m\,s^{-1}}$ with the laser frequency comb (Kotani et al. 2014).

Long-term stability of input illumination to the spectrograph is indispensable for achieving ${\rm 1 \, m\,s^{-1}}$ precision. Variations of observing conditions, such as guiding errors, focus changes, and seeing variations, will change the illuminations at the slit and the point spread function, then limit the RV measurement precision. This is one of the most serious problems for the precise RV method. Recent high-precision RV spectrographs are often fed by optical fibers, for example, HARPS (Mayor et al. 2003), CARMENES (Quirrenbach et al. 2012), and HPF (Mahadevan et al. 2012). Optical fibers can stabilize the slit illuminations, and enable us to locate the spectrograph in a stable place easily.   

However, the output illumination of optical fiber still depends on the input illumination to the fiber. Hence, it is necessary to reduce the correlation by optics, called "scrambling" (Halverson et al. 2015). In addition, the output is also sensitive to variations of the physical conditions of the optical fibers, such as displacements or temperature. This change is caused by the phase and amplitude variations of the finite number of modes interfering at the exit of the fiber, and called "modal noise". 
We need both of the mitigation of modal noise and the  enhancement of scrambling effects by using fiber mode scramblers for high-precision RV measurements. 

The number of propagating modes $M$ in the cylindrical step-index fiber is given by
 
 \begin{eqnarray}
 M = 0.5 \, (\frac{\pi d {\rm NA}}{\lambda})^2
 \end{eqnarray}
 
where $d$ is the diameter of the fiber core, NA is numerical aperture of the fiber,  and $\lambda$ is the wavelength. The modal noise is more significant at NIR wavelengths than visible wavelengths due to less number of propagating modes of light in the optical fiber (Mahadevan et al. 2014).  

Since the targets of exoplanet surveys have until now been mainly Solar-type stars, the RV method has largely been performed at visible wavelengths. Therefore, mode scramblers at visible wavelengths have been relatively well studied. HARPS, for example, achieves an RV precision of better than ${\rm 1 \, m\,s^{-1}}$. On the other hand, studies of mode scramblers at NIR wavelengths (including  how to reduce modal noise) started recently (Roy et al. 2014, Plavchan et al. 2013, Micheau et al. 2012) and there are still many things to be verified experimentally for selecting the best method of a mode scramble.

The number of speckles in the multi-mode fiber end is considered to be the same as $M$ (Daino et al. 1980). If all the speckle patterns are independent, a centroid accuracy of a snapshot would be a degree of $\frac{d}{2\sqrt{M}}$. And then an achievable centroid accuracy might be obtained $\frac{d}{2\sqrt{MN}}$, where N is the number of independent images to be integrated which have different speckle patterns using incident wavefront variations, dynamic scramblers, and wide wavelengths. A worse centroid stability than this accuracy would indicates an imperfect independency of the speckle patterns or the modes in practical cases. 

The modal noise and scrambling effects are measured results at the fiber output surface, and they should be expressed by original physical processes, phase and amplitude changes of the modes in the fiber. The total interacted outcome by a combination of the effects or cascaded mode scramblers would be estimated not by the multiplication of the measured surface results but by the propagation analysis with the physical process parameters. However, it would be a hard work to derive the parameters from the surface measurements. We must also take into account a time variation of the input star light and long exposure of the fiber output. Therefore we have just focused on stabilization degrees of the final output illumination and conducted the same experiments for all the mode scramblers and their combinations including static and dynamic to determine a mode scrambler system providing the highest RV precision under observation conditions of IRD. 

In this paper, we report the results of extensive and systematic laboratory experiments of mode scramblers at NIR wavelengths. We tested three kinds of static mode scramblers [long fibers, nearfield--farfield exchange coupler (NFE coupler), and octagonal fiber] and two kinds of dynamic scramblers and their combinations. To investigate effects of modal noise and the performance of mode scramblers, we measured the output centroid stability to three kinds of disturbances: fiber displacements, movements of entrance pupil, and changes in incident light position at fiber entrance. To make this experiment more realistic, we introduced wavefront errors by a deformable mirror for atmospheric seeing simulation. 

\section{method}
In this section, we describe our experiments in detail. Figure 1 shows the optical system of our experiments. The purpose of our experiments is to decide the best mode scrambler system for IRD. Thus, the optical system simulates the fiber injection module at the nasmyth focus of the Subaru telescope. We used a step-index multi-mode fiber (MMF) with a 60 ${\rm \mu}$m core and an NA of 0.22, named F8950 from OFS Fitel, LLC.. F8950 is used in IRD and its connection to the Subaru telescope in common. We found F8950 has high coupling efficiency, low attenuation, and high scrambling effects by laboratory experiments. \\
The number of modes is about 358 at 1.55um with this fiber. The camera was an Xeva-1.7-640 infrared camera from Xenics.

\begin{figure}[htbp]
\includegraphics[width=9.0cm]{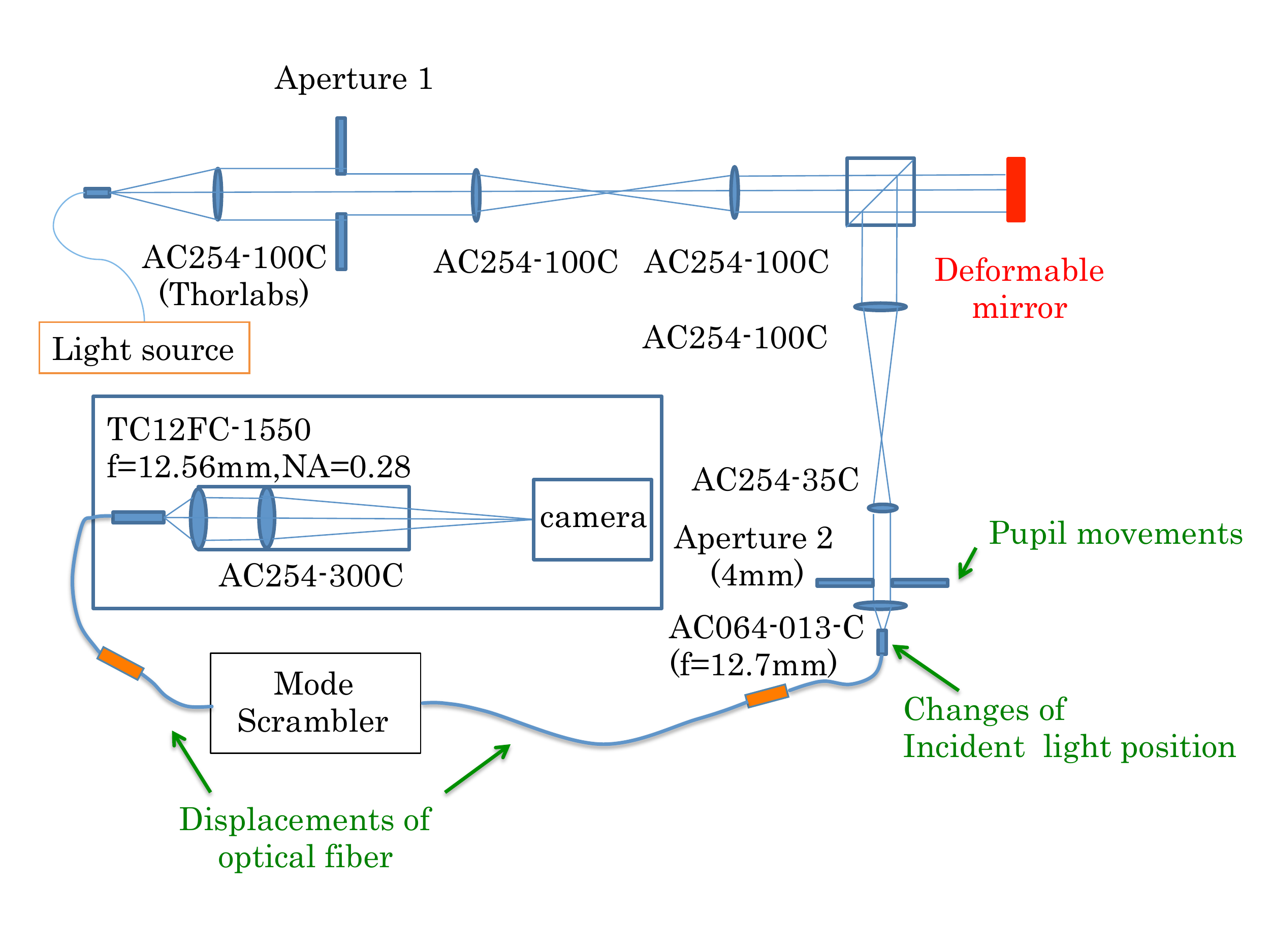}
\caption{The optical system of mode scrambler experiment. All optical fibers of tested mode scramblers had SC connectors, thus they were connected by SC-SC adaptors. Orange rectangles show common SC-SC adaptors to all tested mode scrambler systems.}
\end{figure}

\subsection{Light source}
We used two kinds of light sources: a 1550-nm laser (hereafter, laser) and amplified spontaneous emission (hereafter, ASE) light from an erbium-doped fiber amplifier. ASE has a larger bandwidth; its peak and its full width at half maximum (FWHM) are roughly 1530 nm and 9 nm, respectively. Since a light source with a narrower bandwidth excites a smaller number of propagating modes of light in optical fibers compared with a light source with a wide bandwidth, the centroid of output light for a narrower bandwidth light source is expected to be more unstable. Hence, we used the results with the laser light source to estimate the ultimate RV measurement precision, corresponding to a measurement for almost one absorption line in star's spectrum, and the results of the ASE light source to estimate the precision with all absorption lines in one band; assuming that there are about 500 absorption lines (bandwidth ${\rm \sim}$ 0.02 nm) in each band, the total bandwidth available for RV measurements in one band is 0.02 nm ${\rm \times 500 \sim}$ 10 nm. The wavelength coverage of IRD is Y, J, and H-bands, thus our choice for light sources is conservative; if we can achieve sufficient stability with the ASE light source, we will be able to meet the required stability for the overall instrument.

\subsection{Wavefront error}
 During actual observations, the starlight enters an optical fiber with wavefront errors due to the atmospheric seeing. To simulate this effect, we induced wavefront errors in the light entering the fiber using a deformable mirror. The deformable mirror repeated 6000 frames per 1.5 second. Figure 2 shows the result of the wavefront error production. We could generate various magnitudes of wavefront errors. In this case, achievable accuracy of ${\rm \sim 0.02\,\mu}$m is expected by integrating 6000 independent images (1.5 second integration). This is a theoretical limit of a centroid accuracy with static scrambler (see section 2.3) and the LD light source. This accuracy would be enhanced by dynamic scrambler, because dynamic scrambler produces independent wavefront error from the deformable mirror. However, how much the accuracy would be improved is unknown. 
 
Atmospheric seeing is also expected to have a scrambling effect because the wavefront error would average the energy distribution of excited propagating modes of light in the optical fiber. This suggests that the output centroid from the optical fiber will be unstable when the atmospheric seeing conditions are very good. Therefore, to find a mode scrambler system that has sufficient performance under good seeing conditions, we performed measurements under two magnitudes of wavefront errors that were equivalent to seeing conditions with 0.35" and 0.24"  in FWHM.

 \begin{figure}[htbp]
\includegraphics[width=9.0cm]{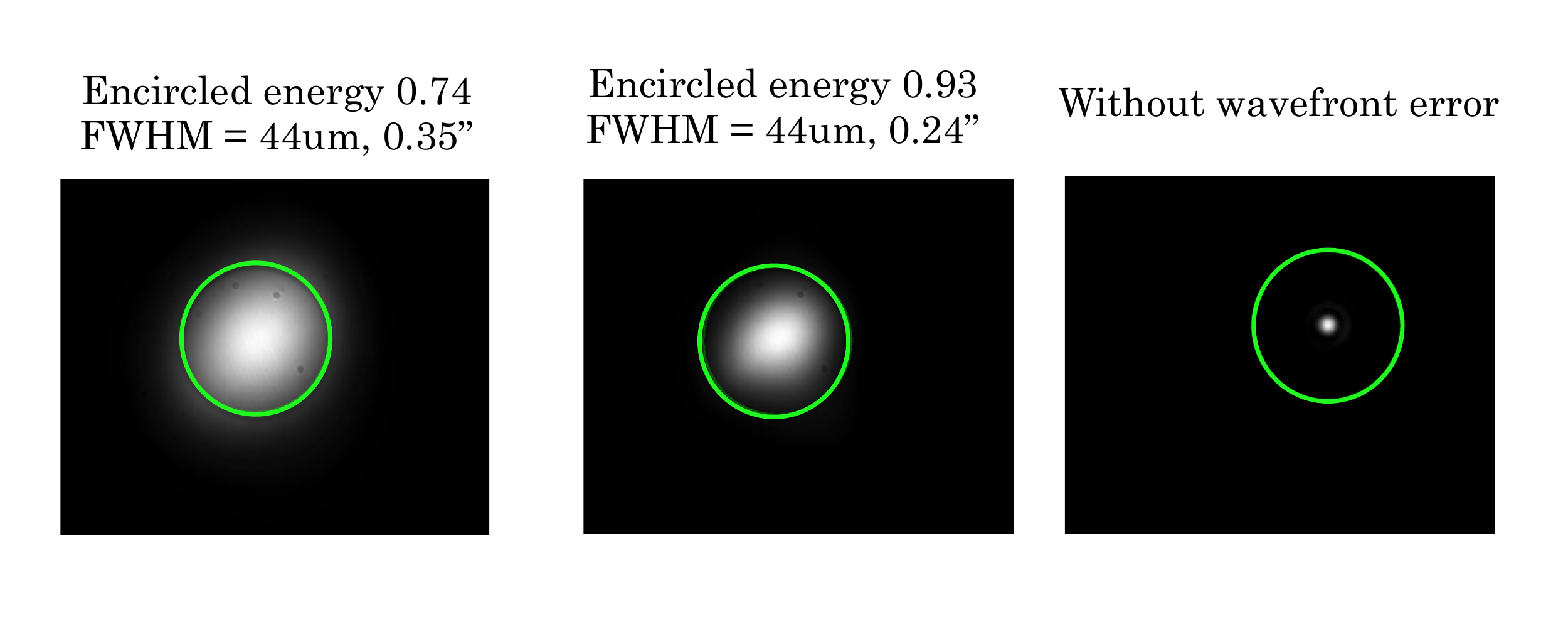}
\caption{Results of seeing simulation by a deformable mirror. The green circles correspond to the 60\, ${\rm \mu}$m core diameter of the optical fiber. The right picture shows the diffraction limited image without wavefront errors. The other two pictures show results of the seeing simulation.}
\end{figure}

\subsection{Mode scrambler}
\subsubsection{Tested mode scramblers}
 We tested various kinds of mode scramblers, including static and dynamic scramblers and their combinations. We considered that there would be some original physical processes to mitigate modal noise and to enhance the scrambling effect. \\
1. Changing phases of existing propagating modes \\
2. Coherence reduction (large phase change) of propagating modes \\
3. Energy redistribution to more propagating modes 

The scrambling effect can be enhanced by 2 and 3. The modal noise can be reduced mainly by 1 but affected also by 2 and 3. A long fiber, for example, was introduced   to obtain 2. However, we conducted the same experiments to the all prepared mode scrambler  systems for systematic understanding, because we have much interest not to separate the effects or the processes but to know stabilization degrees of the final output illumination against to expected disturbances. Details of the tested mode scramblers are given below.
\\

{\bf Long fiber}: An F8950 optical fiber with a length of 65 m is used to transport starlight from the nasmyth focus of the Subaru telescope to the spectrograph. Long fiber can produce large phase shifts between the propagating modes caused by the velocity differences and then enhance the incoherency. In addition, the long fiber winded in a bobbin would redistribute energy to more propagating modes.
We tested F8950 with lengths of 200 m, 243 m, and 509 m. \\

{\bf Octagonal fiber}: Optical fibers with hexagonal or octagonal cores have been suggested to be effective mode scramblers (Avila et al. 2012) like as long fiber. We tested the octagonal core fiber WF 67/125 from Ceramoptec with a length of 10 m. \\

{\bf Nearfield--farfield exchange coupler (NFE coupler)}: A double fiber scrambler (Brown 1990, Hunter \& Ramsey 1992) redistributes energy to more propagating modes, which exchanges the pupil plane and the image plane of the output light from the optical fiber and directs the light to enter the second optical fiber. This type of scrambler is effective at visible wavelengths and also confirmed to be effective at NIR wavelengths (Halverson et al. 2015). We exchanged the nearfield and farfield patterns by a coupler shown in Figure 3. We call this type of scrambler as "NFE coupler" in this paper.\\

\begin{figure}[htbp]
\includegraphics[width=9.0cm]{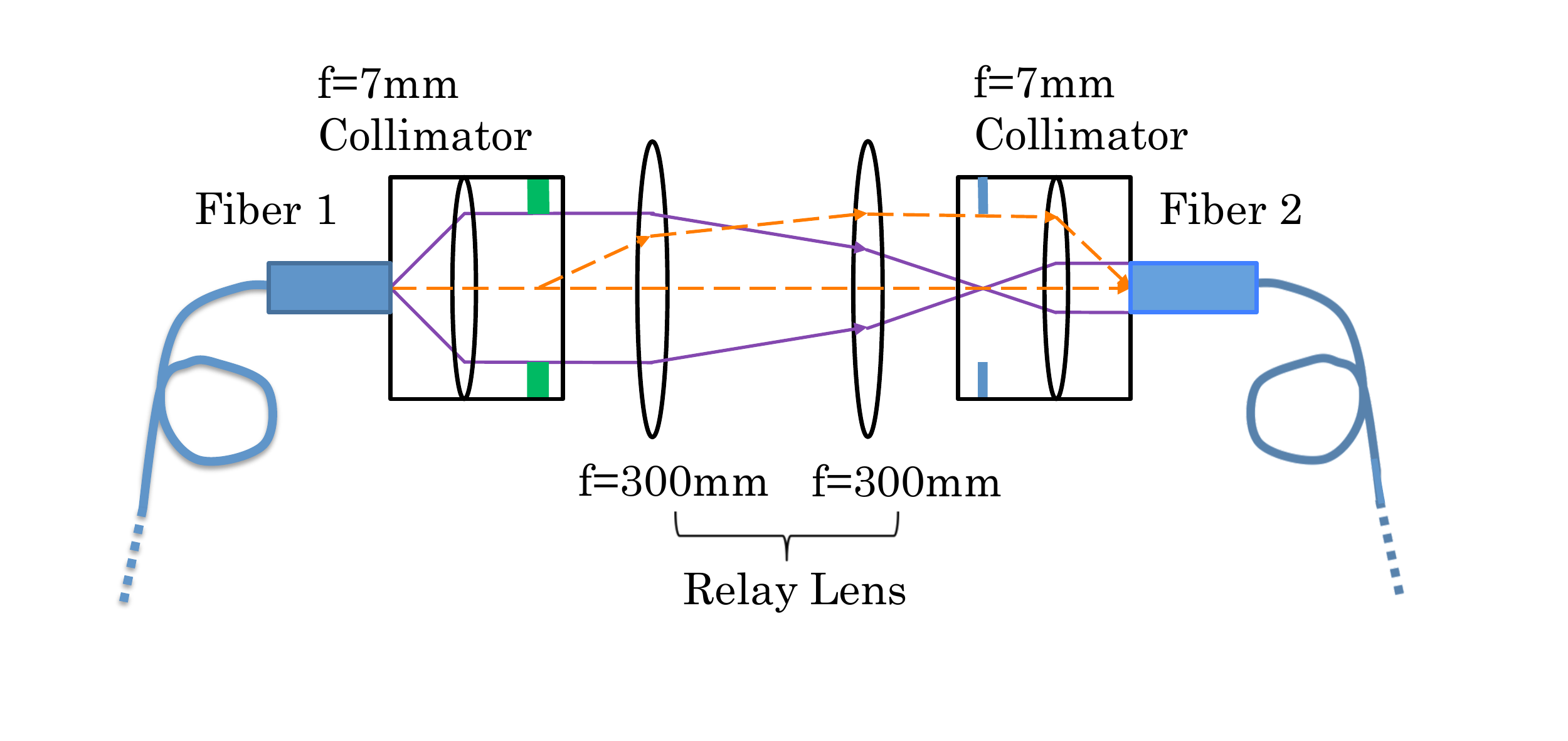}
\caption{An outline of our NFE coupler. We used  RC02FC-P01 for reflective collimator, and AC254-300-C for relay lenses from Thorlabs. We defined pupil (shown by green bars) and formed the pupil image onto the end of second fiber. Solid (purple) and dotted (orange) lines show that NFE coupler exchanges the pupil plane and the image plane of the output from the fiber 1.}
\end{figure}

{\bf Dynamic scrambler}: This type of scramblers, which move optical fibers dynamically, have been suggested to be effective mode scramblers at NIR wavelengths (McCoy et al. 2012, Mahadevan et al. 2014). This kind of mode scrambler produces phase shifts of the light in the optical fiber by placing stress on optical fibers. We tested two kinds of dynamic scrambler. The first one is a ``bending" optical fiber using a seesaw shaker for mixing of chemical liquids. Figure 4 shows an outline of our bending scrambler. The bended optical fibers are F8950 and WF67/125 with lengths of 10 m. Fibers are looped ten times with a radius of 20 cm and fitted to a seesaw shaker. Thus, there are 20 bending points with bend radius of  25 mm that will be moved simultaneously by the seesaw shaker. We chose bend radius of 25 mm to prolong fiber lifetimes (long-term bend radius of F8950 shown by the manufacturer is 17 mm). The second type of dynamic scrambler is ``twisting" scrambler made by Giga Concept Inc.. This scrambler twists the optical fiber by two electric motors. The twisted optical fiber is a 60-${\rm \mu}$m-core fiber with a length of roughly 1 m from LEONI. 

\begin{figure}[htbp]
\includegraphics[width=8.0cm]{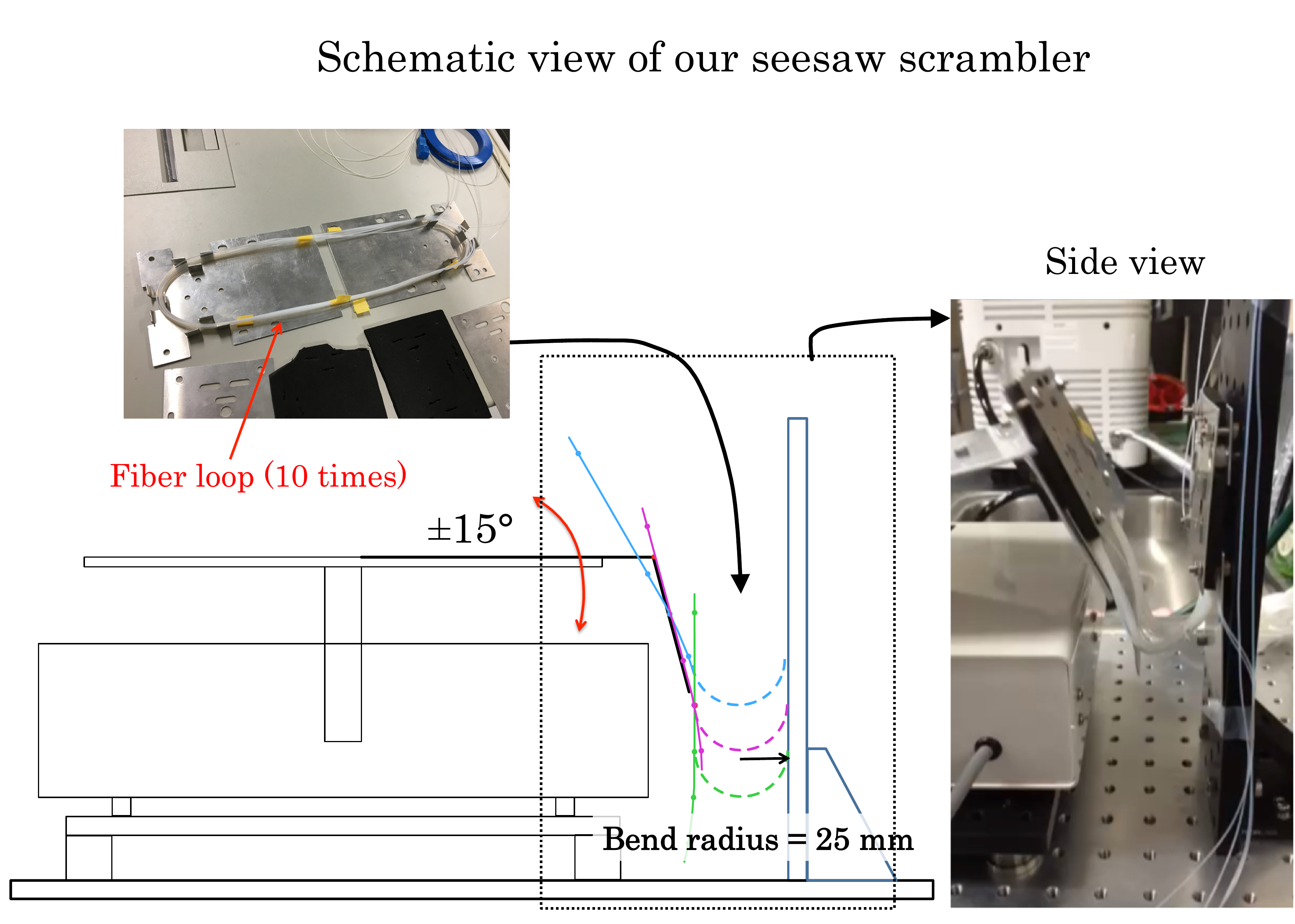}
\caption{Outline and picture of our bending scrambler. An optical fiber with a length of 10 m was looped 10 times with a radius of 10 cm, and fitted to a seesaw shaker along the diameter. The seesaw shaker induces bends with radius of 25 mm in opposite sides of the fiber loop. Thus, total 20 positions of the optical fiber were bended simultaneously. The period of the seesaw shaker movements was roughly 1 second.}
\end{figure}

\subsubsection{Efficiency}
Table 1 lists the efficiency of tested mode scramblers at 1550 nm.  Octagonal fibers  have lower efficiencies compared with the other mode scramblers due to connection. This is because our octagonal fiber has the different core diameter from F8950. The NFE coupler also has lower efficiency. One of the causes is the reflections by lenses and collimators. The values on the Table 1 are measured efficiencies at the time of experiments. We note that the efficiency of NFE coupler would be improved ($\sim$90\%) by optimizing the optics.
\begin{table}[h]
\begin{center}
\caption{{\normalsize Tested mode scramblers and efficiency (at 1550 nm)}}
\begin{tabular}{ccc}
\hline \hline \smallskip
Scrambler & Abbreviation & Efficiency   \tabularnewline
 \hline \smallskip
NFE coupler & NFEC & 80\%  \tabularnewline
\hline \smallskip 
Long fiber (e.g.\ 509 m) & 509m & 90\%/500m \tabularnewline 
\hline \smallskip 
Octagonal fiber & Oct & 80\%  \tabularnewline 
\hline \smallskip 
Bending scrambler (dynamic) & Bending & ${\rm \geq 98\%}$ \tabularnewline
\hline \smallskip 
Twisting scrambler (dynamic) & Twisting & ${\rm \geq 98\%}$ \tabularnewline 
\end{tabular}
\end{center}
\end{table}

\subsection{Disturbance}
 The input illumination to the spectrograph must be stable in the long term, though various observation conditions will change temporally. Therefore, to investigate the effects of these varying observing conditions, we applied three kinds of disturbances to the optical system and measured the shifts of the output centroids.

\subsubsection{Displacements of optical fibers}
 A telescope moves during observations to track stars, and the optical fibers will also move along with the telescope. Furthermore, the temperature of the optical fibers will also change. Hence, the physical conditions of the optical fibers will vary during observations, and this variation will cause modal noise. To simulate these variations, we displaced the optical fibers before the mode scrambler system 25 cm every 10 seconds, as shown in Figure 5. We displaced the optical fibers eight times in one measurement, and the integration time in each fiber position is 6 seconds (0.5 second per frame). We also displaced the optical fiber after the mode scramble system to investigate whether the modal noise can arise again after the mode scrambler system. 
 
\begin{center}
\begin{figure}[htbp]
\includegraphics[width=9.0cm]{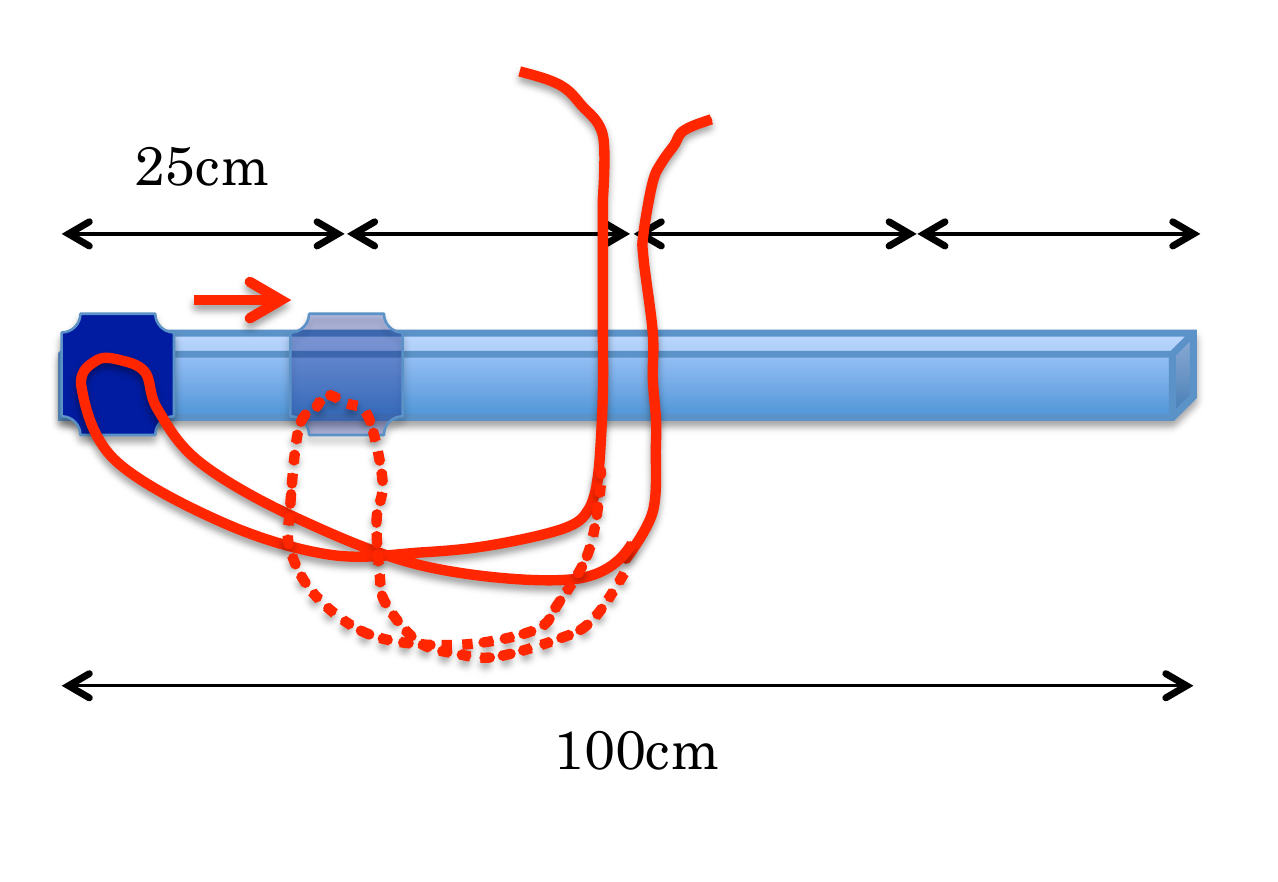}
\caption{The method of fiber displacement experiments. We fixed the optical fibers to a slider on an optical rail and displaced the slider 25 cm every 10 seconds.  We displaced the optical fiber eight times (a round-trip of the optical rail with a length of 1 meter). }
\end{figure}
\end{center}

\subsubsection{Pupil movements}
 During observations, due to imperfect optical axis alignment, the entrance pupil will move slightly as the telescope rotates. Although it is not known how much the entrance pupil moves at the Subaru telescope, we moved aperture 2, shown in Figure 1, and measured shifts of the output centroids to estimate the effect of the pupil movements. We moved aperture 2 in the horizontal direction up to ${\rm \pm\, 400\, \mu}$m with a ${\rm \pm \,100  \,\mu}$m step. The diameter of aperture 2 is 4 mm and thus the effect of up to 10\% pupil movement can be estimated by this experiment. The integration time was 15 seconds (1 second per frame, a total of 15 frames).

\subsubsection{Changes of incident light positions to the optical fiber}
 Due to variations in atmospheric dispersion corrector effects, the incidence position of the incident light on the optical fiber will change during actual observations. To simulate this effect, we moved the fiber end (shown in Figure 1) and changed the incident light position by optical manipulators. The position could be changed both up and down and left and right relative to the optical axis. We moved the fiber end up to 20 ${\rm \mu}$m with a 5 ${\rm \mu}$m step and measured the shifts of the output centroids. The integration time was 15 seconds (1 second per frame, a total of 15 frames).
 
 \subsection{Intensity distributions of the fiber end}
  Intensity distributions of the fiber end is also important. Shifts of output centroid directly degrades RV measurement precision as above, and remaining structures on a fiber top-hat profile also affect the RV measurement. The remaining structures would change the slit illumination when the coupling optics between the fiber end and the slit is distorted by a temperature change. The change of the remaining structure even without the centroid shift, would distort the point spread function of the spectrograph and would affect the RV. To investigate this issue, we have just calculated azimuthally averaged power spectrum of the remaining structures to each mode scrambler system, which indicates the spatial-frequency resolved components of the variance of the remaining structure.

\subsection{FFP measurement}
 Variations in the farfield pattern (FFP) also affects the RV measurement precision. However, it is difficult to predict how much the FFP of incident light would change due to variations in guiding, focus, and atmospheric seeing during long-term observations. In addition, the effect of the output FFP variation on the RV measurement precision also depends on the aberration of the spectroscopic optical systems. Thus, it is difficult to estimate the effects of the output FFP variation from the optical fiber against the RV measurement precision (St\"{u}rmer et al. 2014).
 Focal ratio degradation (FRD) is also a serious problem. FRD means that the F-number of the output light from optical fibers is smaller than that of the input light. In the case of IRD, the star light enters the optical fiber with a numerical aperture (NA) of 0.15 at the nasmyth focus, and the output light that spreads outside an NA of 0.15 is cut off by an optical aperture in the spectrograph. Therefore, the NA of the output light from the optical fiber has a great influence on the efficiency of IRD. Thus, we also measured the output FFP distribution and measured the magnitude of FRD induced by each mode scrambler.
 We put the fiber end 12.7 mm away from the detector instead of the NFP measurement optical system (Figure 1), and measured the distribution of the output FFP.

\section{Results}
 We tested 24 mode scrambler systems. In the case of IRD, a ${\rm 0.01 \,\mu m}$ centroid shift of the input light to the spectrograph will cause an ${\rm \sim\, 1\, m/s}$ RV measurement error. Thus, we used this value (${\rm 0.01 \,\mu m}$) as a standard for evaluations of tested mode scrambler systems.
  Figure 6 shows an example of our experimental results. We took an image of the MMF end and an image of the single mode fiber (SMF) end simultaneously. Since the SMF has only one propagating mode of the light, the output light from the SMF has no modal noise. Hence, we used the image of the SMF as the positional standard of the centroid calculation for the MMF image to remove any effects due to shifts of the entire optical system. 
 
\begin{figure}[htbp]
\includegraphics[width=9.0cm]{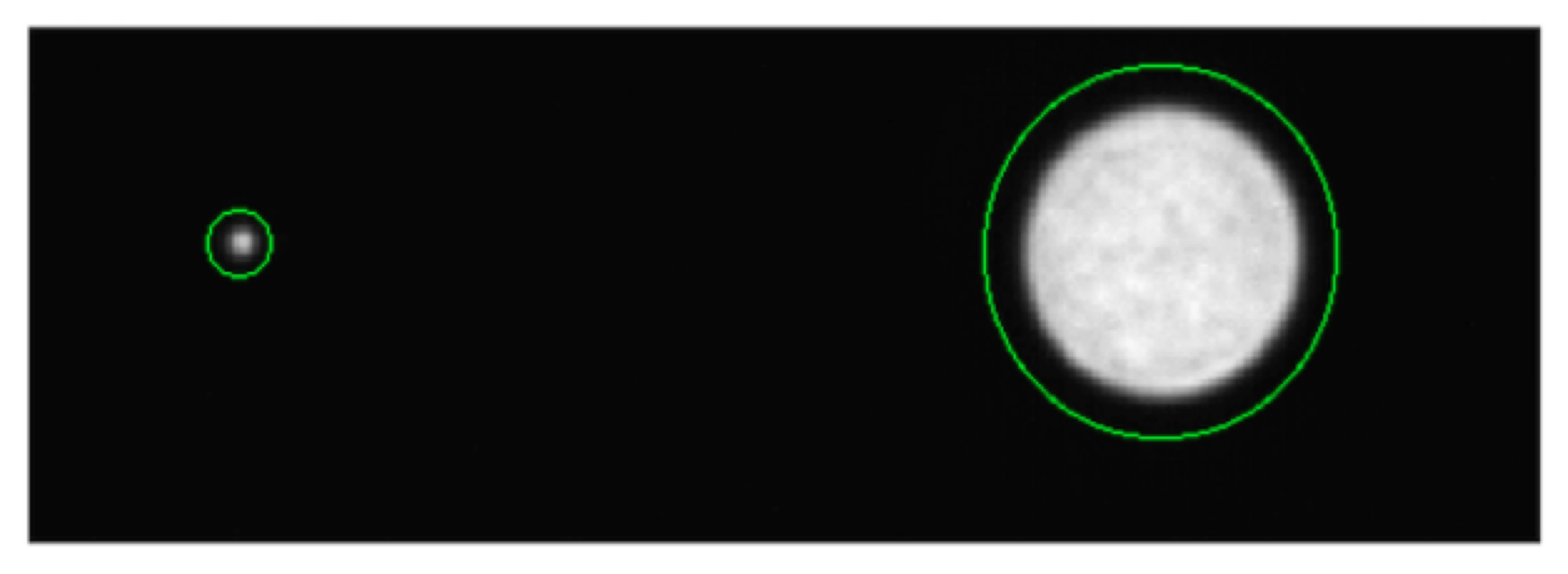}
\caption{Example of our experiments. The large circle on the right is an image of the MMF end, while the small circle to the left is an image of the SMF end. We used the centroid of the SMF image as a standard for the centroid calculation for the MMF image. The green circles show the regions used for centroid calculation for the MMF and SMF images. }
\end{figure}

\subsection{Displacements of optical fibers}
 The purpose of this experiment is to investigate which mode scrambler system is effective against variations in an optical fiber's physical conditions (e.g.\ displacement, temperature). We calculated the performance of each mode scrambler system as follows: first, we calculated the centroid of images with 6-second integration (except for 2 seconds before and after each fiber displacement). We investigated centroid shifts during a measurement (180 seconds), and we found 2 seconds were enough time for output speckle pattern to settle into a new equilibrium. We displaced the optical fiber eight times during one measurement and obtained nine images per one measurement. The largest centroid displacement in nine images, we call this "maximum variation width", will be used as a performance indicator of tested mode scrambler system.  
 
\begin{figure*}[htbp]
\begin{center}
\includegraphics[width=17cm]{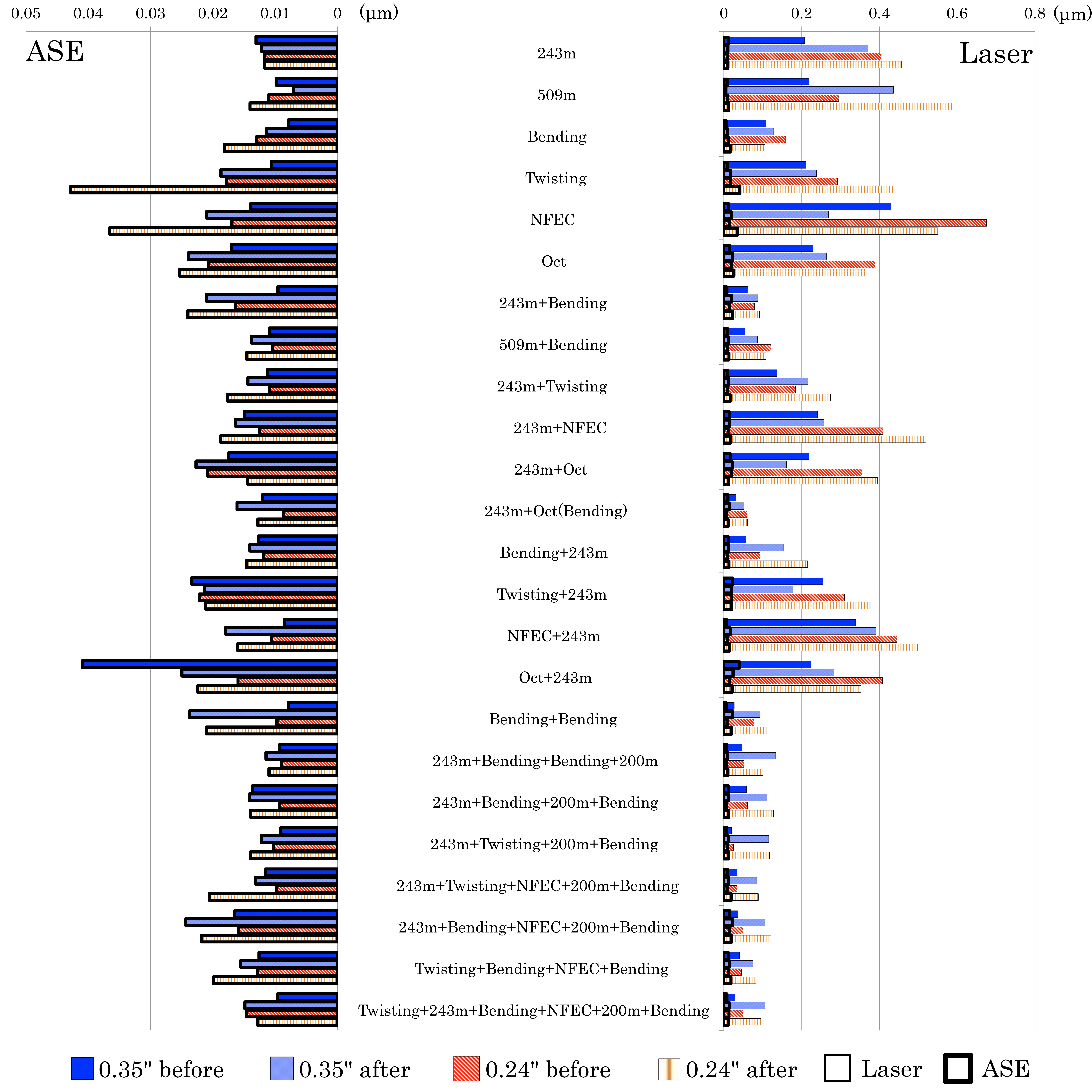}
\caption{Result of fiber displacement experiments. The left figure shows the result with the ASE light source and the right figure shows the result with the laser light source. The ``before" and ``after" labels indicate the positions of the displaced optical fiber against the mode scrambler system. To compare these results, the results with the ASE light source are included in the right figure by black outlined boxes overlaid on the results for the laser light source. }
\end{center}
\end{figure*}
Figure 7 summarizes all our results. Note that we used approximately 2 meters of F8950 optical fiber to connect the mode scramblers, which is not noted in Figure 7 and other figures in this section. The results with the laser light source show that dynamic scramblers (especially the bending type) are effective against fiber displacements, like as preceding study (Baudrand \& Walker, 2001). We confirmed that agitating many positions of the optical fibers with a large amplitude is effective against modal noise in the configuration of IRD. We also found that the performance of dynamic scramblers is enhanced by using a long fiber before the dynamic scrambler. This might be a result that a long fiber increases the relative phase difference between the propagating modes and a dynamic scrambler varies the phase of each propagating mode. These various and large phase changes will scramble the phase changes caused by variations in an optical fiber's physical condition, which will lead to average out a modal noise with a certain integration time.  

The maximum variation width of all the measured mode scrambler systems were substantially smaller for the ASE light source. Thus, we confirmed that the effect of the physical condition variations of optical fibers against the RV measurement precision can be mitigated greatly by increasing the bandwidth for RV measurement (corresponding to using more absorption lines). In addition, the maximum variation width of each mode scrambler was roughly the same value. We speculate that this is because the maximum variation widths for the ASE light source induced by the fiber displacement were too small and below the measuring limit of our experiment. To confirm the threshold of this experiment, we conducted the same measurements without fiber displacements. As a result, the maximum variation width for the ASE light source without fiber displacements was ${\rm \sim 0.01 \, \mu m}$. Therefore, the results with the ASE light source were limited by the laboratory setup and environment. Differences in the measurement conditions (e.g.\ temperature) or subtle differences in the precise fiber displacements might be the cause of the differences between the results for each mode scrambler system. Altogether, we did not measure any significant differences between the mode scrambler systems with the ASE light source, we confirmed that we would be able to achieve an output centroid stability with a level of ${\rm \leq \sim 0.02 \, \mu m}$ against fiber displacement by using all the absorption lines in a band.

\subsection{Pupil movements}
\ Figure 8 shows the result of pupil movements experiments, demonstrating that the variations of the entrance pupil  affect the output intensity distribution by changing the energy distribution of the excited propagating modes in the optical fiber. To evaluate and compare the performance of each mode scrambler system we applied linear-function fitting (Avila \& Singh, 2008) to the result shown in Figure 8. We assumed that the shift of the output centroid ({\bf  y}) is proportional to the magnitude of the pupil movement ({\bf x}), and the factor of proportionality ({\bf a}) can be calculated by a linear-function fitting ({\bf  y}={\bf  ax}) for each result. Hence, we can consider that a mode scrambler system with small {\bf a} is effective against pupil movements. Figure 9 shows the results of fitting (with a factor of proportionality {\bf a} for each mode scrambler system).

In the case of the laser light source, it is found that a combination of a long fiber and a dynamic scrambler is effective against pupil movements, as with the experiments for fiber displacements. Coherence reduction and changing phases of propagating modes for a long time also have a scrambling effect. Especially, the combination of a long fiber, a dynamic scrambler, and an NFE coupler showed a high performance. This result is consistent with other studies (e.g.  Roy et al. 2014).

On the other hand, an NFE coupler is effective with the ASE light source. The optical fiber has a high scrambling effect on the image plane but only a small scrambling effect on the pupil plane; thus exchanging the image plane and the pupil plane by an NFE coupler would be effective against pupil movements. Furthermore, the performance of an NFE coupler is enhanced when it is combined with a long fiber or a dynamic scrambler. 

We confirmed that broadening the bandwidth of the light source is not effective against pupil movements, in contrast to fiber displacement. Therefore, the wavelength dependence of the output centroid shift caused by pupil movements would be small. 
 The most stable mode scrambler system is ``Twisting + 243m + Bending + NFEC + 200m + Bending", and can reduce shifts of the output centroid below ${\rm 0.01 \,\mu m}$ against 10\% pupil movements.

\begin{figure*}[htbp]
\begin{center}
\includegraphics[width=16cm]{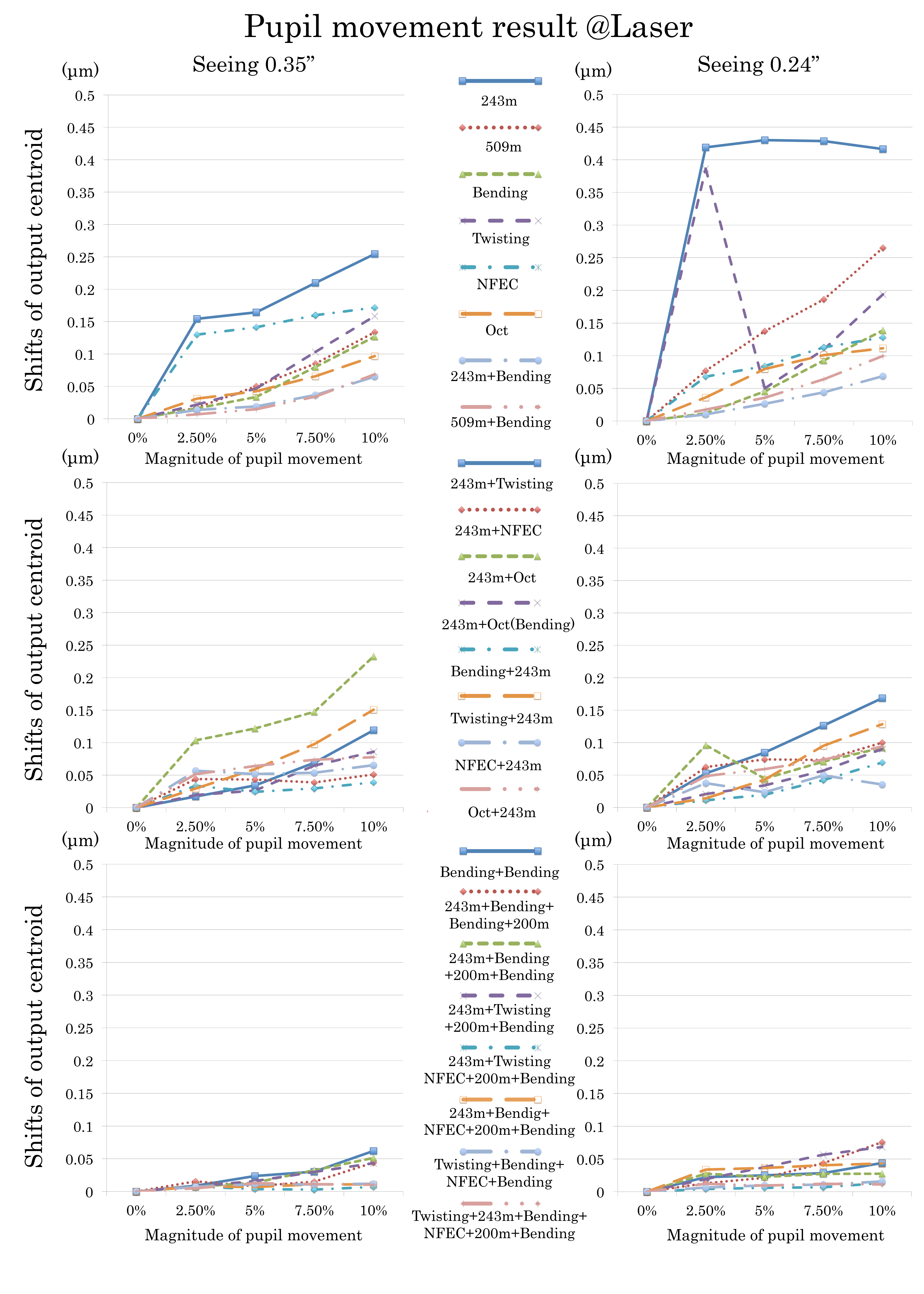}
\end{center}
{\bf Figure 8(a).}\ Results of pupil movement experiments with the Laser light source. The left figures show the result for 0.35" seeing and the right figures show the results for 0.24" seeing. 
\end{figure*}

\begin{figure*}[htbp]
\begin{center}
\includegraphics[width=16cm]{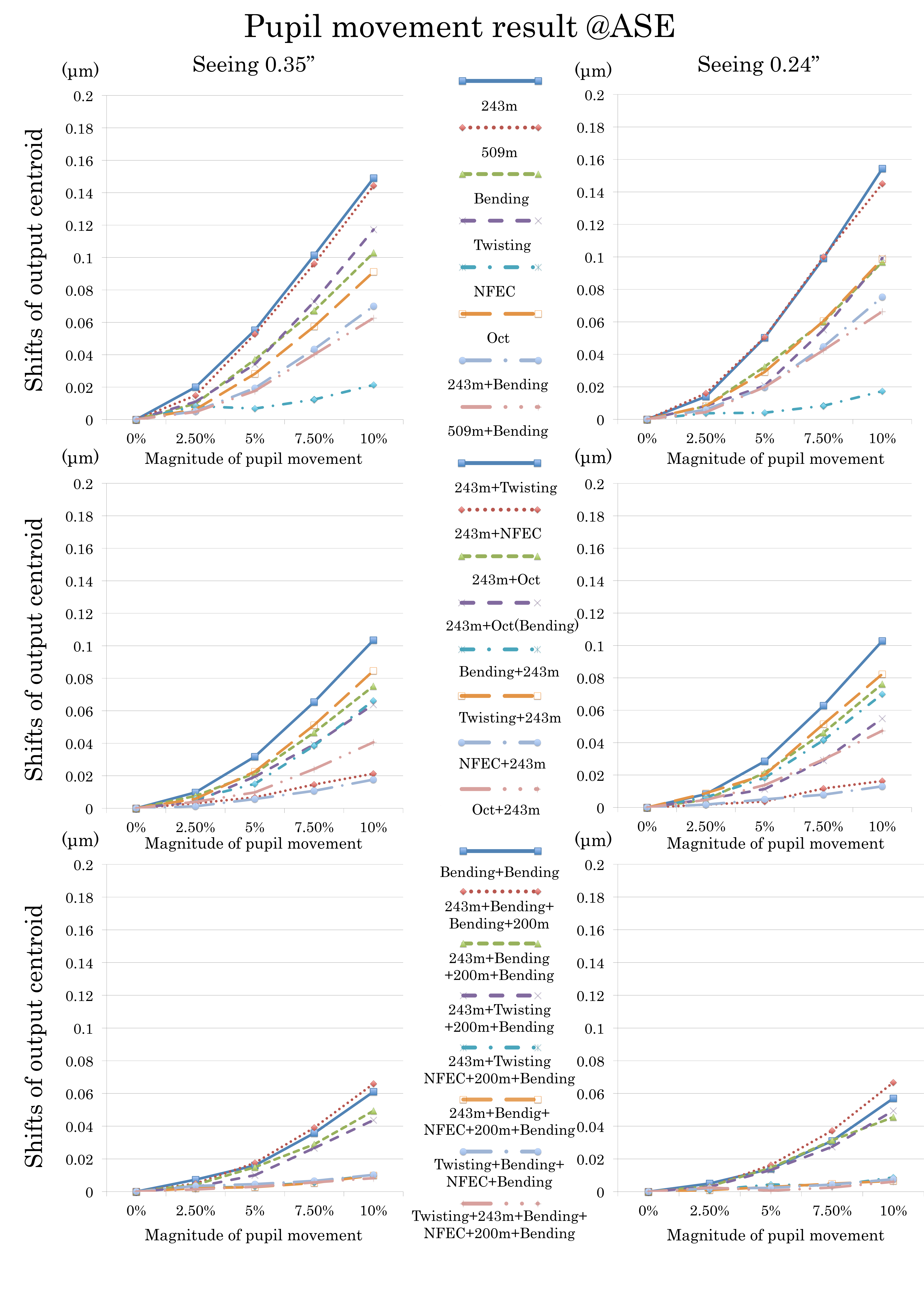}
\end{center}
{\bf Figure 8(b).}\ Results of pupil movement experiments with the ASE light source. The left figures show the result for 0.35" seeing and the right figures show the results for 0.24" seeing. 
\end{figure*}
\setcounter{figure}{8}

\begin{figure*}[htbp]
\begin{center}
\includegraphics[width=18cm]{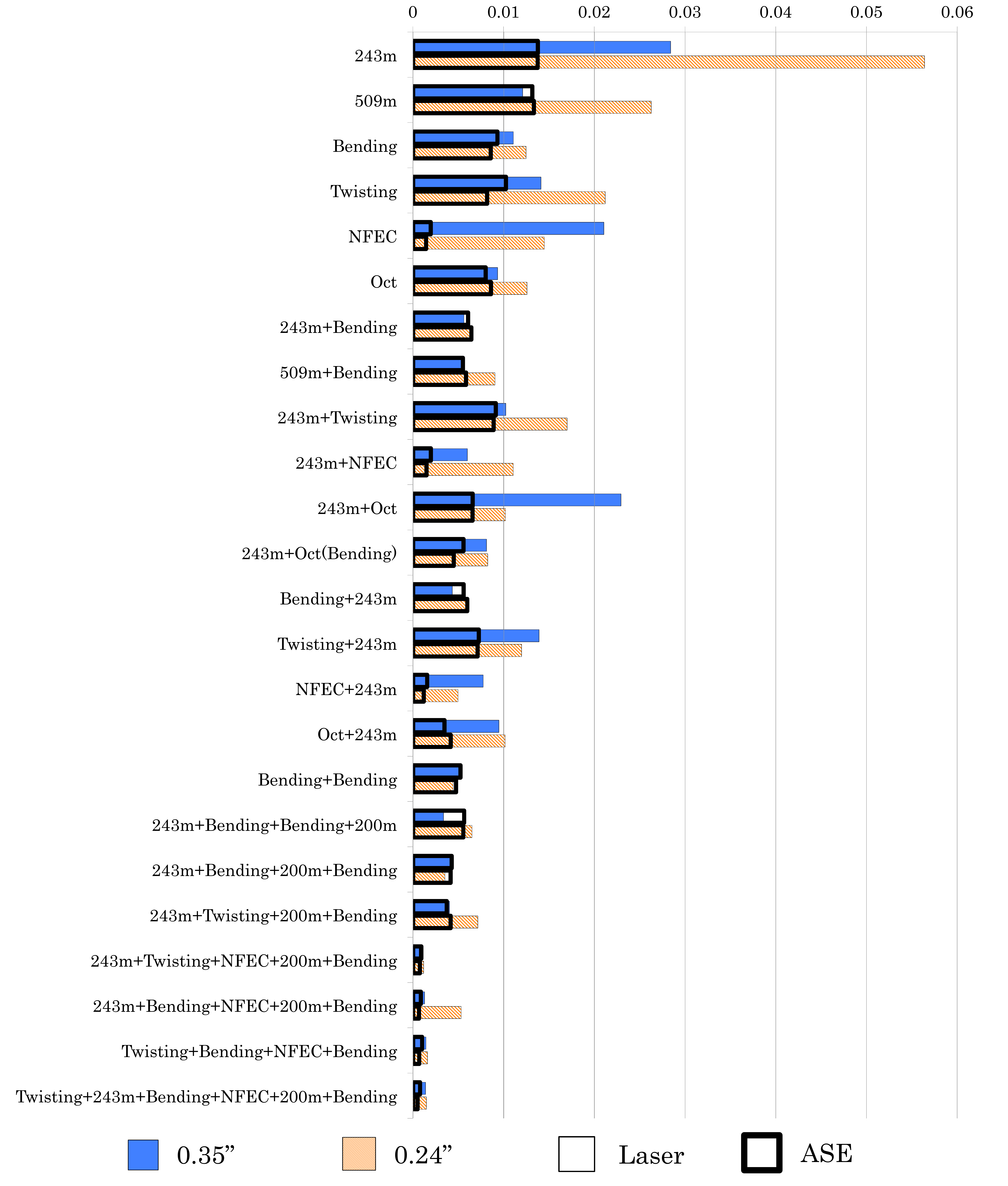}
\caption{Linear-function fitting to the results shown in Figure 8. This figure shows the factor of proportionality of each mode scrambler system. The factor of proportionality with the ASE light source is shown by black outlined boxes over the results with the laser light source for easy comparison.}
\end{center}
\end{figure*}

 \begin{figure*}[htbp]
 \begin{center}
\includegraphics[width=16cm]{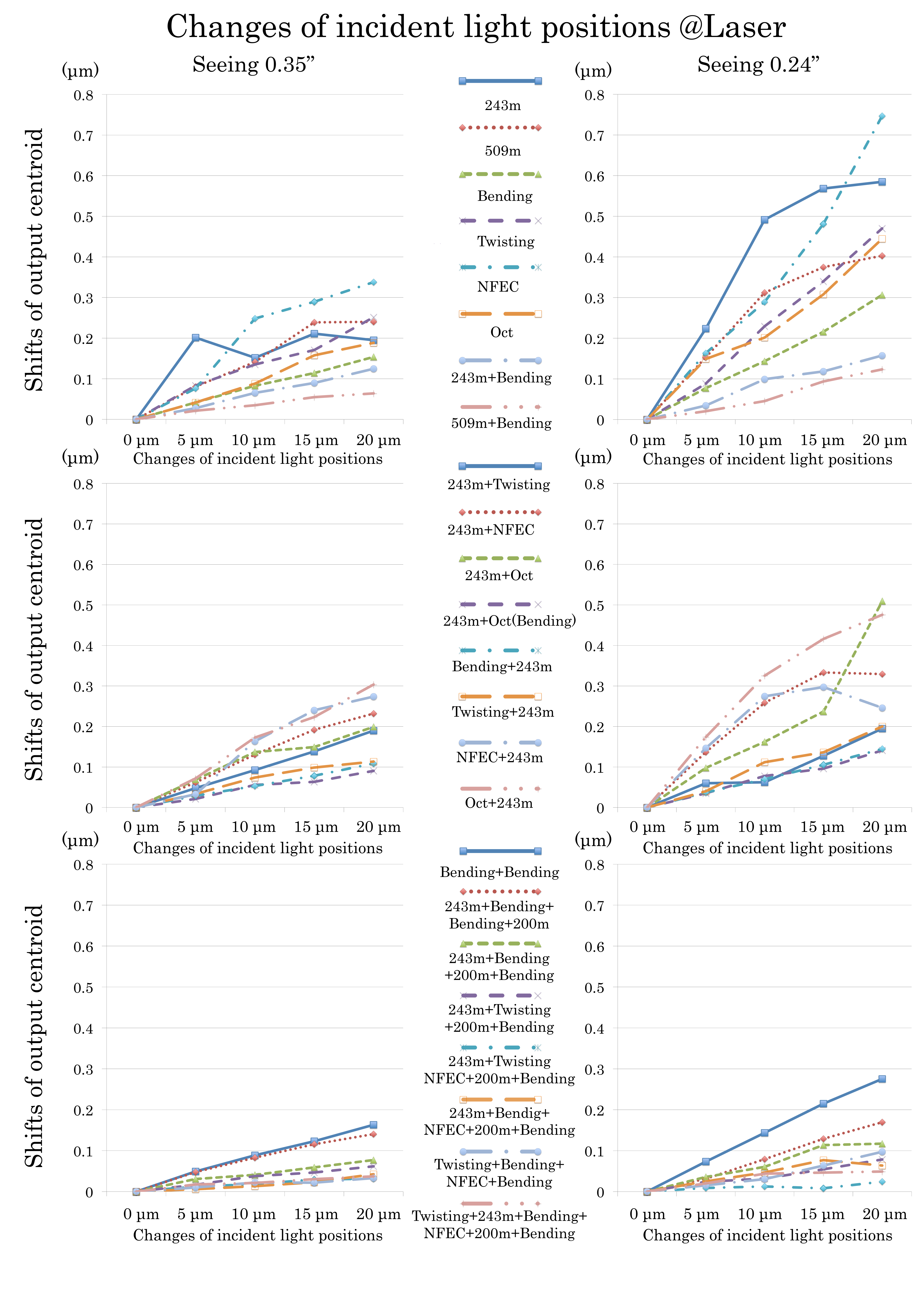}
\end{center}
{\bf Figure 10(a).}\ Results of incident light position change experiments with the Laser light source.  The left figures show the result for 0.35" seeing and the right figures show the results for 0.24" seeing. 
\end{figure*}

\begin{figure*}[htbp]
\begin{center}
\includegraphics[width=16cm]{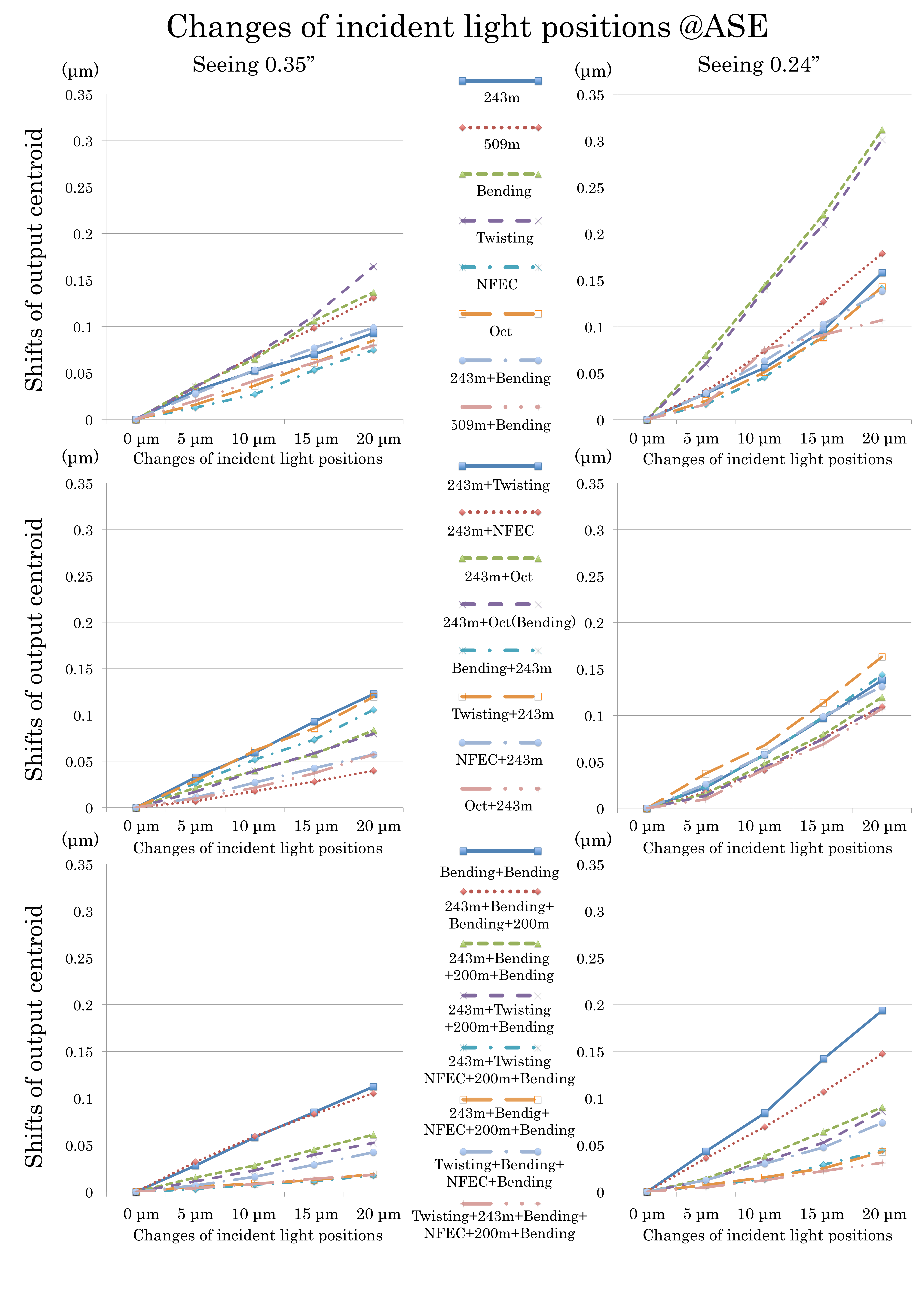}
\end{center}
{\bf Figure 10(b).}\ Results of incident light position change experiments with the ASE light source. The left figures show the result for 0.35" seeing and the right figures show the results for 0.24" seeing. 
\end{figure*}

\setcounter{figure}{10}

\begin{figure*}[htbp]
\begin{center}
\includegraphics[width=18cm]{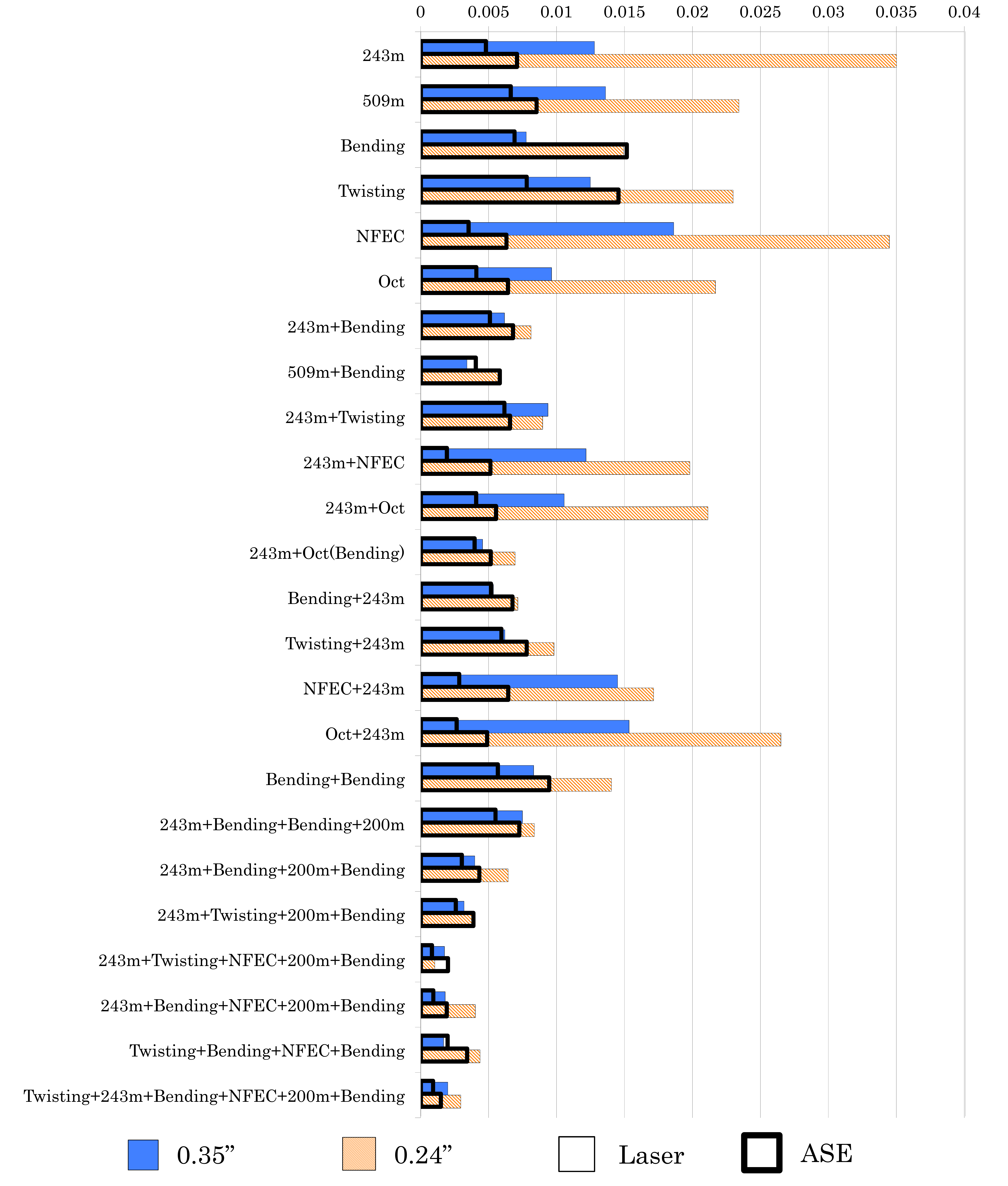}
\caption{Linear-function fitting to the results shown in Figure 10. This figure shows the factor of proportionality for each mode scrambler system. }
\end{center}
\end{figure*}

 \begin{figure*}[htbp]
 \begin{center}
\includegraphics[width=16cm]{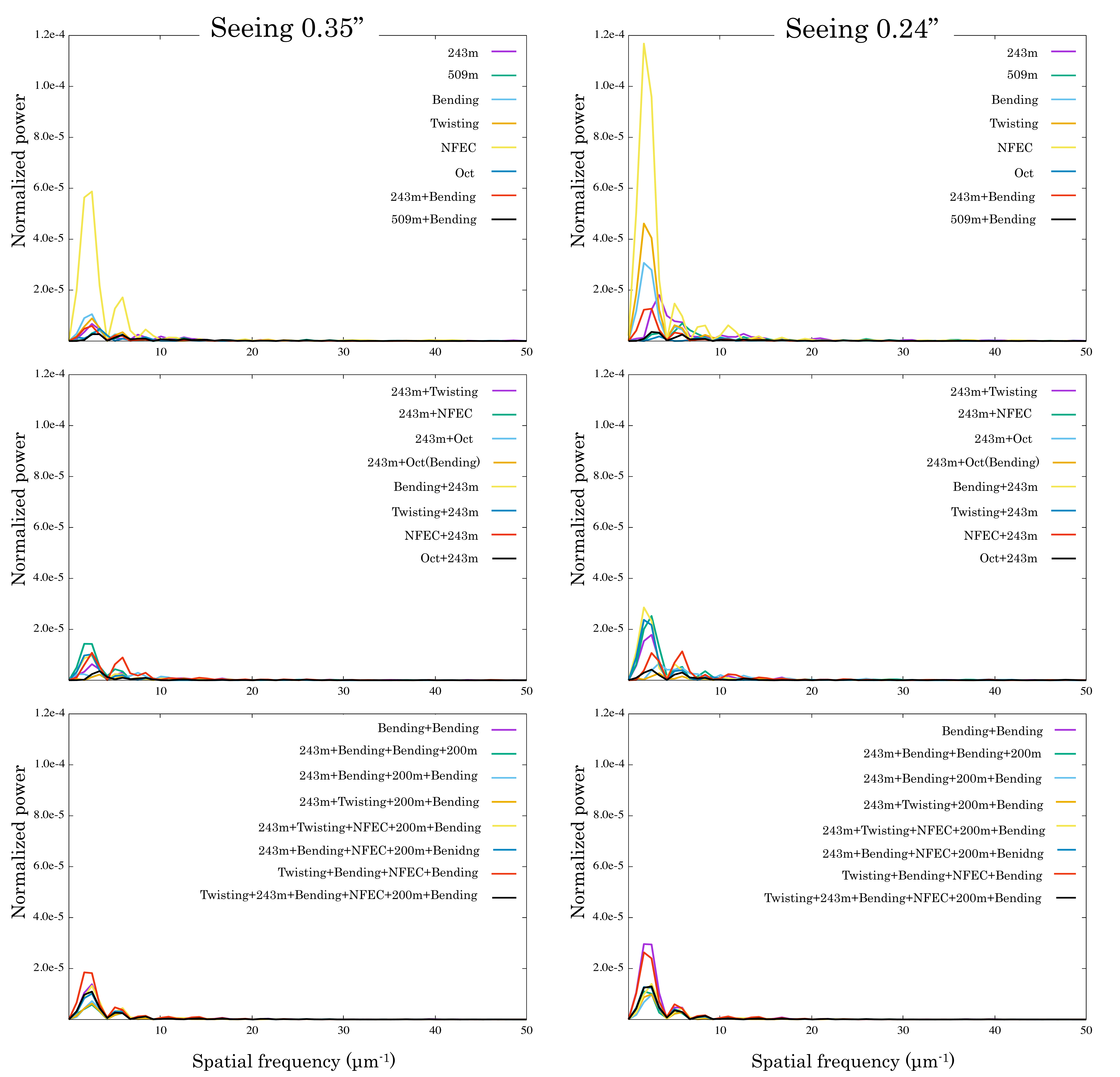}
\end{center}
{\bf Figure 12(a).}\ Power spectra of azimuthally averaged intensity distributions after subtracting fitted top-hat profiles with the LD light source. Intensity distributions of scrambler systems which had long fibers were flattened well.
\end{figure*}

\begin{figure*}[htbp]
\begin{center}
\includegraphics[width=16cm]{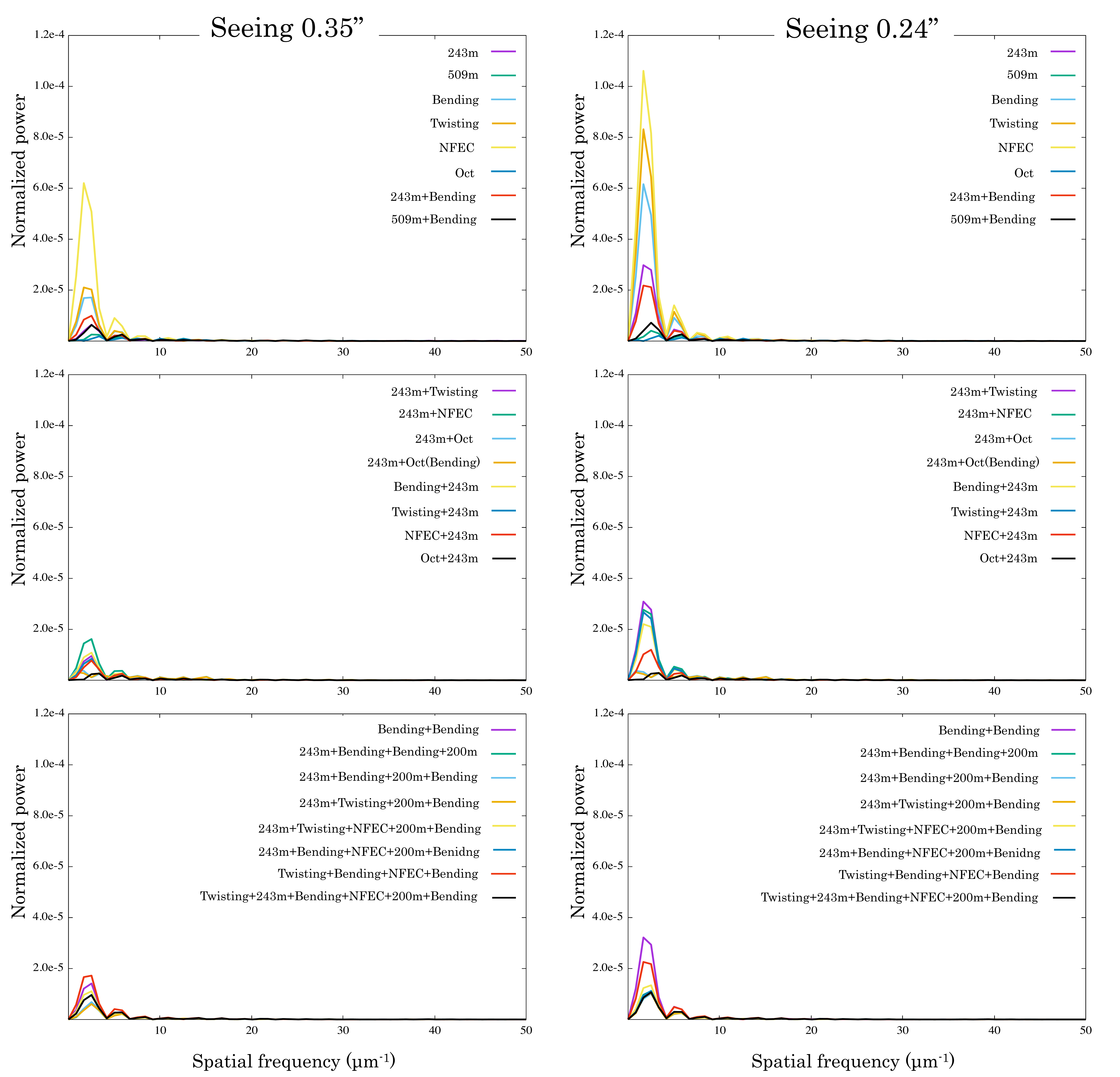}
\end{center}
{\bf Figure 12(b).}\ Power spectra of azimuthally averaged intensity distributions after subtracting fitted top-hat profiles with the ASE light source. 
\end{figure*}

\subsection{Changes of incident light positions to the optical fiber}
 Figure 10 shows the results of the experiment. The results indicate that changes in the incident light positions, as well as pupil movement, induce centroid shifts of the output centroid by changing the energy distribution of the excited propagating modes in the optical fibers. As with the experiments on pupil movements, we applied linear-function fitting to the results shown in Figure 10 for a concise evaluation. Figure 11 shows the results of fitting, that is, the factors of proportionality of the relationship between the shifts of the output centroid and the magnitude of the incident light position change. 
 
The NFE coupler was not effective with the laser light source, though it exhibited a strong effect with the ASE light source. The performance of the NFE coupler was enhanced by being combined with a long fiber and a dynamic scrambler.
 Broadening the bandwidth of the light source is not effective against changes in the incident light position, in contrast to the case of fiber displacement. This feature was also found for the case of pupil movements. Thus, we confirmed that it is difficult to mitigate the output variations caused by variations in the incident light conditions by broadening the bandwidth of the light source. Furthermore,  in general, mode scrambler systems that are effective against pupil movements are also effective against changes in incident light position. This would be because both disturbances induce changes of the energy distribution of the excited propagating modes in the optical fibers. 
 The most effective mode scrambler system, as with the case of pupil movements, was "Twisting + 243m + Bending + NFEC + 200m + Bending". This mode scrambler system can reduce shifts of the output centroid below ${\rm  0.01\, \mu m}$ against 5${\rm \, \mu m}$ shifts of the incident light position.

\subsection{Intensity distributions of the fiber end}
We investigated that how much tested mode scramblers could flatten the intensity distributions of the fiber end. Figure 12 shows power spectra of azimuthally averaged intensity distributions of the fiber end. We note that we subtracted a fitted top-hat profile from the intensity distribution before calculation of the power spectrum for concise comparison. 

We found that the intensity distribution of "NFE coupler only" had relatively large remaining structures. We speculate that the original far field pattern coupled to the fiber end affects the remaining structure, because there was only short length optical fiber with the "NFE coupler only" configuration. The remaining structures in this case were flattened by combining with other mode scramblers, especially long fiber. This shows that coherence reduction of propagating modes is effective to flatten the output intensity distribution. We also found that long fiber can flatten the output intensity distributions even with good seeing conditions. Long fiber can feed stable and flat output intensity distributions against variations of the seeing condition.  \\

\setcounter{figure}{12}

\subsection{FFP measurement}
 Figure 13 shows the NA and the half width at half maximum (hereafter, HWHM) of the tested mode scrambler systems, and Figure 14 shows the efficiency of the FFP, which were calculated by the below calculation,
 \begin{eqnarray}
 \rm{\frac{Count\, \,within\, \,NA 0.15}{Total\,\,count}}  \nonumber
 \end{eqnarray}
 
 Note that the FFP was measured with the ASE light source because there were too many interference fringes in the FFP with the laser light source.
 
The spread angles of the output light from the mode scrambler systems that include a NFE coupler were relatively large, and therefore the FFP efficiencies of such systems is relatively low. The FFP efficiencies of other mode scrambler systems were roughly 80\%. It is found that combining various mode scramblers does not affect the FFP efficiency. As a result, we concluded that the differences in FFP efficiency caused by differences in the mode scrambler system are small (a few percent at most). Thus FRD is not a serious problem in the tested mode scramblers.
\\
\section{Conclusion}
To study planet formation and detect Earth-like planets in the HZ around M dwarfs, we have developed IRD, which is an NIR high-precision Doppler spectrograph for the Subaru telescope. The target RV measurement precision of IRD is ${\rm \sim1 \, m\,s^{-1}}$. In order to achieve this RV measurement precision, it is necessary to stabilize the incident light to the spectrograph by using a mode scrambler system. However, the most effective mode scrambler system at NIR wavelengths is still unknown. Thus, we conducted systematic research of various mode scramblers and their combinations including static and dynamic with the same experiments and evaluation methods, the stability of output illumination of the fiber, under realistic conditions, such as atmosphere seeing simulations by a deformable mirror. \\
 We applied three kinds of disturbances to the optical system to simulate changes in observational conditions, and measured the centroid shifts of the output intensity distribution. As a result, we obtained the following findings. \\
{\bf Fiber displacements}:\ \ The combination of a long fiber and a dynamic scrambler is effective. Broadening the bandwidth of the light source significantly reduces shifts in the output centroid. Using more absorption lines for RV measurement is an effective way to reduce the effect of fiber displacement.\\
{\bf Pupil movements \& changes of incident light position}:\ \ The combination of "Twisting + 243m + Bending + NFEC + 200m + Bending" is effective. Broadening the bandwidth of the light source does not reduce shifts of the output centroid significantly, in contrast to the case with fiber displacements.

The loss caused by FRD is ~20\% and is roughly the same value for all measured mode scrambler systems.
 Considering these facts and the efficiency of each mode scrambler, we concluded that the best scrambler system candidates for IRD are \\
(A) Twisting + 243m + Bending + NFEC + 200m + Bending \\
(B) Twisting + 243m + Bending + 200m + Bending 

\ We note that all fibers in the above mode scrambler systems are F8950 from OFS Fitel LLC, a Furukawa electric company, except for a fiber from Leoni in the Twisting scrambler.  (twisted fiber have the same core-diameter with F8950). System (A) was the most powerful mode scrambler system in our experiments. This system can reduce the shifts of an output centroid below ${\rm  0.01\, \mu m}$ against 10\% pupil movements or a ${\rm 5\,\mu m}$ change of incident light position. However, the efficiency for light utilization of system (A) is relatively low because the system includes an NFE coupler. In addition, the actual magnitudes of the observing condition variations are unknown. Therefore, another mode scrambler system with a lower performance might be able to achieve sufficient stability. Hence, we added system (B) as an alternative efficient mode scrambler system. The definitive choice for the mode scrambler system will be determined based on engineering observations, which have been conducted from 2017. After the final decision, we will splice all mode scramblers to enhance overall throughput.  \\

This project is supported by the JSPS Grant-in-Aid (Nos. 15H02063 and 22000005) and a NINS Astrobiology Center Grant-in-Aid.

 \begin{figure*}[htbp]
\begin{center}
\includegraphics[width=16cm]{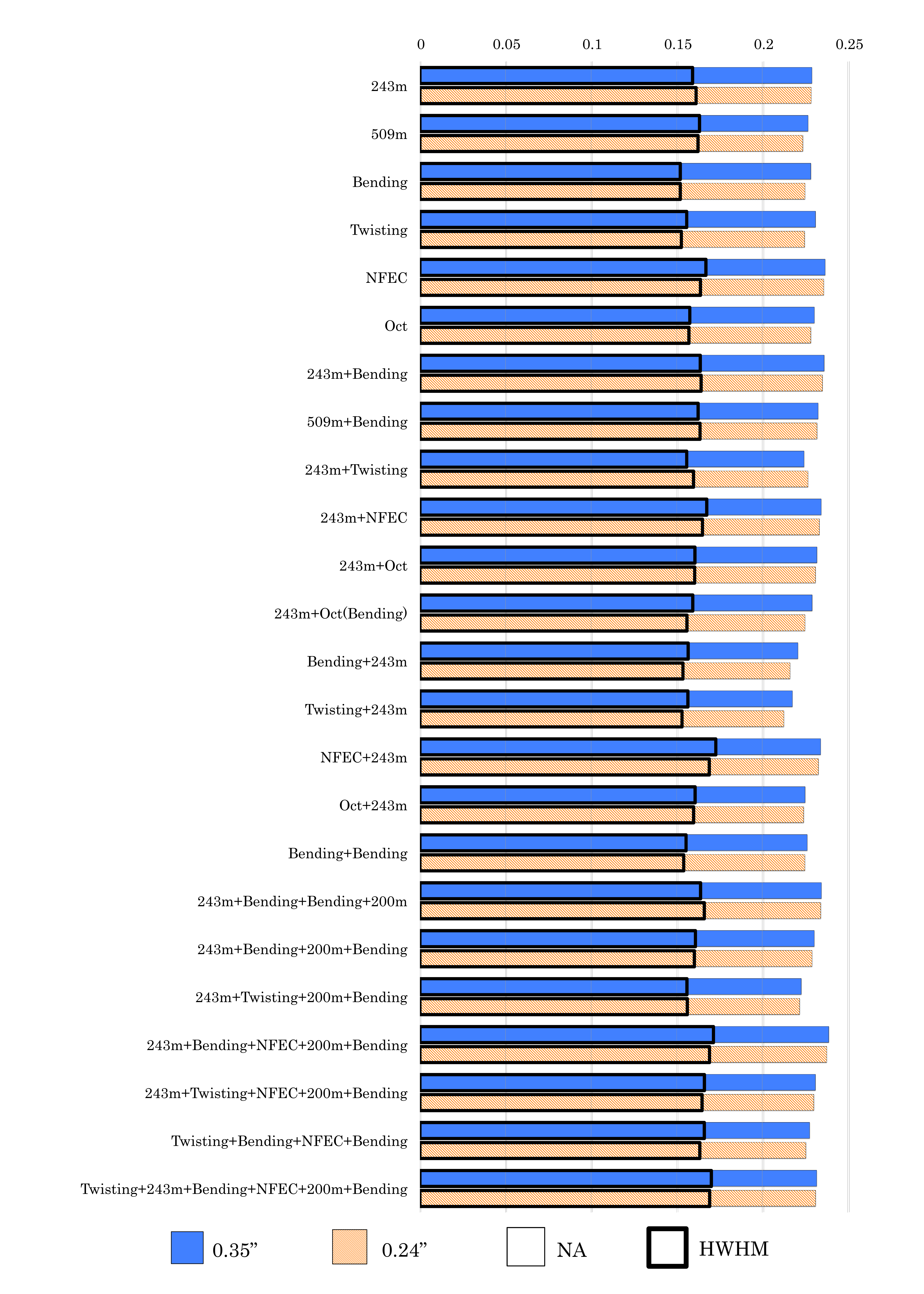}
\caption{NA and  HWHM of tested mode scrambler FFP distributions. HWHM is shown by black outlined boxes over the values of NA. The unit of this figure is the sine of the angle.}
\end{center}
\end{figure*}
\begin{figure*}[htbp]
\begin{center}
\includegraphics[width=16cm]{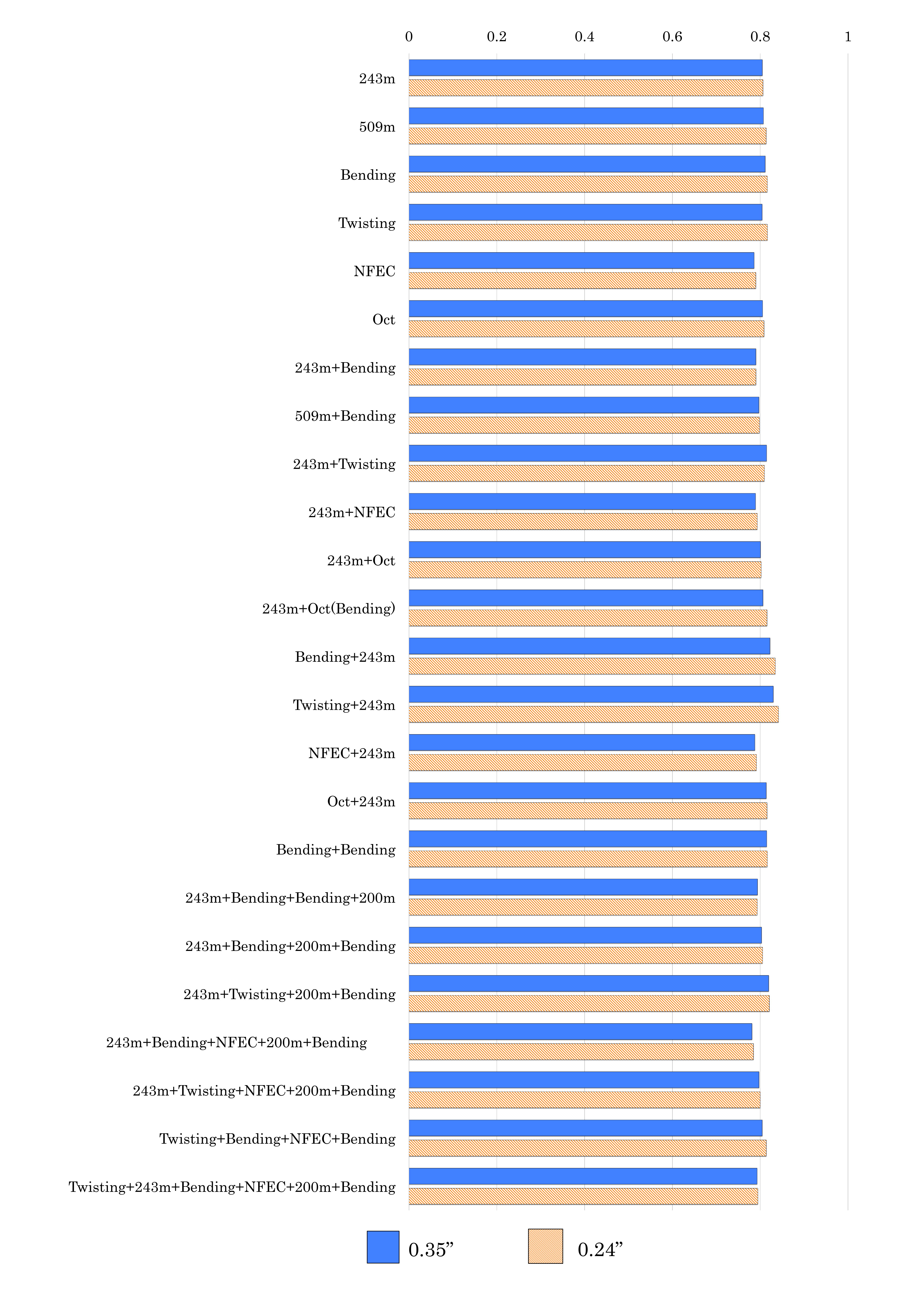}
\caption{FFP efficiency of tested mode scrambler. }
\end{center}
\end{figure*}

\begin{center}
{\scshape REFERENCE}
\end{center}

\hangindent=20pt
\noindent
Avila, G., 2012, Proc. SPIE 8446, Ground-based and Airborne Instrumentation for Astronomy IV, 84469L 

\hangindent=20pt
\noindent
Avila, G., \& Singh, P., 2008, Proc. SPIE 7018, Advanced Optical and Mechanical Technologies in Telescopes and Instrumentation, 70184W 

\hangindent=20pt
\noindent
Brown, T. M., 1990, ASPC, 8, 335B 

\hangindent=20pt
\noindent
Daino, B. et al., 1980, Journal of Modern Optics, 27, 1151

\hangindent=20pt
\noindent
Halverson, S.  et al., 2015, ApJ, 806, 61H 

\hangindent=20pt
\noindent
Huang, S., 1959, PASP, 71,421H 

\hangindent=20pt
\noindent
Hunter, T. R., \& Ramsey, L. W. 1992, PASP, 104, 1244H 

\hangindent=20pt
\noindent
Kokubo, T. et al., 2016, Proc. SPIE 9912, Advances in Optical and Mechanical Technologies for Telescopes and Instrumentation II, 99121R 

\hangindent=20pt
\noindent
Kotani, T. et al., 2014, Proc. SPIE 9147, Ground-based and Airborne Instrumentation for Astronomy V, 914714 

\hangindent=20pt
\noindent
Kopparapu, R. K. et al., 2013, ApJ, 765, 131K 

\hangindent=20pt
\noindent
Mahadevan, S. et al., 2012, Proc. SPIE 8446, Ground-based and Airborne Instrumentation for Astronomy IV, 84461S 

\hangindent=20pt
\noindent
Mahadevan, S. et al., 2014, ApJ, 786, 18M 

\hangindent=20pt
\noindent
Mayor, M. et al., 2003, Msngr, 114, 20M 

\hangindent=20pt
\noindent
McCoy, K. S. et al., 2012, Proc. SPIE 8446, Ground-based and Airborne Instrumentation for Astronomy IV, 84468J 

\hangindent=20pt
\noindent
Micheau, Y. et al., 2012, Proc. SPIE 8446, Ground-based and Airborne Instrumentation for Astronomy IV, 84462R 

\hangindent=20pt
\noindent
Plavchan, P. P. et al., 2013, Proc. SPIE 8864, Techniques and Instrumentation for Detection of Exoplanets VI, 88641J 

\hangindent=20pt
\noindent
Quirrenbach, A. et al., 2012, Proc. SPIE 8446, Ground-based and Airborne Instrumentation for Astronomy IV, 84460R 

\hangindent=20pt
\noindent
Roy, Arpita. et al., 2014, Proc. SPIE 9147, Ground-based and Airborne Instrumentation for Astronomy V, 91476B 

\hangindent=20pt
\noindent
St\"{u}rmer, J. et al., 2014, Proc. SPIE 9151, Advances in Optical and Mechanical Technologies for Telescopes and Instrumentation, 915152

\end{document}